\documentclass{osa-article}
\usepackage{units}
\usepackage{ulem}
\journal{osajournal}

\articletype{Research Article}

\begin{document}

\title{Polarization Spectroscopy of High-Order Harmonic Generation in Gallium Arsenide}

\author{Shatha Kaassamani\authormark{1}, Thierry Auguste\authormark{1}, Nicolas Tancogne-Dejean\authormark{2}, Xu Liu\authormark{1,3}, Willem Boutu\authormark{1}, Hamed Merdji\authormark{1}, and David Gauthier\authormark{1,*}}
\address{\authormark{1}Université Paris-Saclay, CEA, CNRS, LIDYL, 91191 Gif-sur-Yvette, France\\}
\address{\authormark{2}Max Planck Institute for the Structure and Dynamics of Matter and Center for Free-Electron Laser Science, Luruper Chaussee 149, 22761 Hamburg, Germany\\}
\address{\authormark{3}Imagine Optic, 18, rue Charles de Gaulle, 91400 ORSAY, France\\}

\email{\authormark{*}david.gauthier@cea.fr}


\begin{abstract}
An interesting property of high harmonic generation in solids is its laser polarization dependent nature which in turn provides information about the crystal and band structure of the generation medium. Here we report on the linear polarization dependence of high-order harmonic generation from a gallium arsenide crystal. Interestingly, we observe a significant evolution of the anisotropic response of above bandgap harmonics as a function of the laser intensity. We attribute this change to fundamental microscopic effects of the emission process comprising a competition between intraband and interband dynamics. This intensity dependence of the anisotropic nature of the generation process offers the possibility to drive and control the electron current along preferred directions of the crystal, and could serve as a switching technique in an integrated all-solid-state petahertz optoelectronic device. 
\end{abstract}

\section{Introduction}
Since its first observation in 2010 \cite{ghimire_observation_2010}, continuous and extensive research has been carried out to understand the fundamental physical processes underlying high harmonic generation (HHG) in semiconductors and dielectrics. HHG was investigated in a number of crystals including semiconductors and insulators \cite{ghimire_observation_2010,you_anisotropic_2017,luu_generation_2018,kim_generation_2017}, as well as 2D materials \cite{yoshikawa_high-harmonic_2017,taucer_nonperturbative_2017,liu_high-harmonic_2017,Langer2018}. An interesting characteristic of HHG in crystals is the strong dependence of the generation process on the structural and optical properties of the medium, and its response to the laser parameters, such as intensity and polarization. This dependence has been exploited as a spectroscopic tool to access, for instance, the electron dynamics and retrieve the band structure \cite{vampa_all-optical_2015,Lanin2017,Tancogne-Dejean2017,Itatani2018,zhao_theory2019,Klemke2019,polar_spectro_zno_2019,lightwave_tomography2020,tempointerfero_peixiang2020}. The polarization-dependent nature of the HHG process has been explained from a real-space picture \cite{you_anisotropic_2017,real-space_recollision2019} as a result of the crystal lattice structure, or correspondingly from a momentum-space domain representation which involves the crystal band structure \cite{vampa_all-optical_2015,Lanin2017}. As a direct consequence of the band configurations in momentum space, two main mechanisms are involved in the emission of photons for a given harmonic: the interband polarization and the nonlinear intraband current. The former necessitates the recombination of an electron–hole pair between two energy bands, while the latter originates from the nonlinear oscillation of electrons in non-parabolic band curvatures. These two mechanisms yield a generation efficiency that reflects the crystal or band structure. For instance, Lanin et al. \cite{lanin_high-order_2019} have mapped the anisotropic band dispersion of a ZnSe crystal through the polarization and intensity-dependent below bandgap harmonics, dominated by intraband currents. However, for above bandgap harmonics both interband and intraband mechanisms can contribute to the harmonic emission, and a competition might arise between them especially when electrons explore higher energy regions of the band structure. Here, we report on our study of high harmonic generation in gallium arsenide (GaAs) both in transmission and reflection geometries. We investigate the dependence of above bandgap harmonics on the laser polarization for different laser intensities in the nonperturbative regime. We find out that the polarization dependence of low order harmonics changes with increasing laser intensity, in such a way that the maximum harmonic yield reverses from one crystal axis to another. In the transmission geometry, we characterize the laser propagation through the crystal and prove that nonlinear propagation effects have  minimal impact on the change in the HHG anisotropy maps measured. This is further supported by measurements carried out in reflection geometry, where the same intensity-dependent polarization mapping is obtained. We conclude that our observations portray pure electron dynamics involved in interband and intraband mechanisms in GaAs.

\section{Experimental Setup and Details}
The HHG experiment in transmission geometry (t-HHG) is performed by employing an all-fiber mid-IR laser (NOVAE Brevity \cite{NOVAE}), delivering 82 fs, 9 nJ laser pulses at a repetition rate of 18.66 MHz centered at a wavelength of 2.1 µm. The beam is focused by a 3 cm focal length lens onto a 500 µm thick gallium arsenide (GaAs) crystal having a (100) cut. The laser intensity and polarization are varied by a half-wave plate and a polarizer. To take into account propagation effects in the GaAs sample, the output power as well as the beam size near the exit surface are measured to estimate the effective intensities responsible for the HHG. This is detailed in section 3.2. For the HHG experiment in reflection geometry (r-HHG), we employ an OPCPA (Optical Parametric Chirped Pulse Amplification, Starzz from Fastlite) laser system providing higher energy pulses, reaching 10 µJ, at a central wavelength of 2.4 µm. The repetition rate is 100 kHz and the pulse duration is 75 fs. The laser pulses are focused by a 7.5 cm focal length lens at ${45^{\circ}}$ angle of incidence onto the 500 µm thick GaAs (100) crystal. The estimated beam size at the focus on the sample is around 40 µm in the vertical direction. The reported effective intensities in the reflection configuration are estimated taking into account the incidence angle and the Fresnel transmission coefficient for an s-polarized state. For both experiments, t-HHG and r-HHG, the harmonics generated are collected and focused by a 10 cm focal length lens onto a CCD camera and a spectrometer. The spectral selection of each harmonic is done using band pass filters centered at each harmonic wavelength. Given that GaAs is a narrow direct bandgap semiconductor with a bandgap energy of 1.4 eV, all generated harmonics are above bandgap harmonics. The direct bandgap emissions are measured in the spectrum from GaAs around a wavelength of 880 nm. In addition, another less intense fluorescence signal at 450 nm is measured, which corresponds to higher direct transitions in the conduction bands at the ${\Gamma}$ point. These two fluorescent signals are filtered out spectrally or discriminated spatially on the camera.

\section{Results}
\subsection{HHG in Transmission Geometry} 
\begin{figure}[htbp]
\centering\includegraphics[width=12cm]{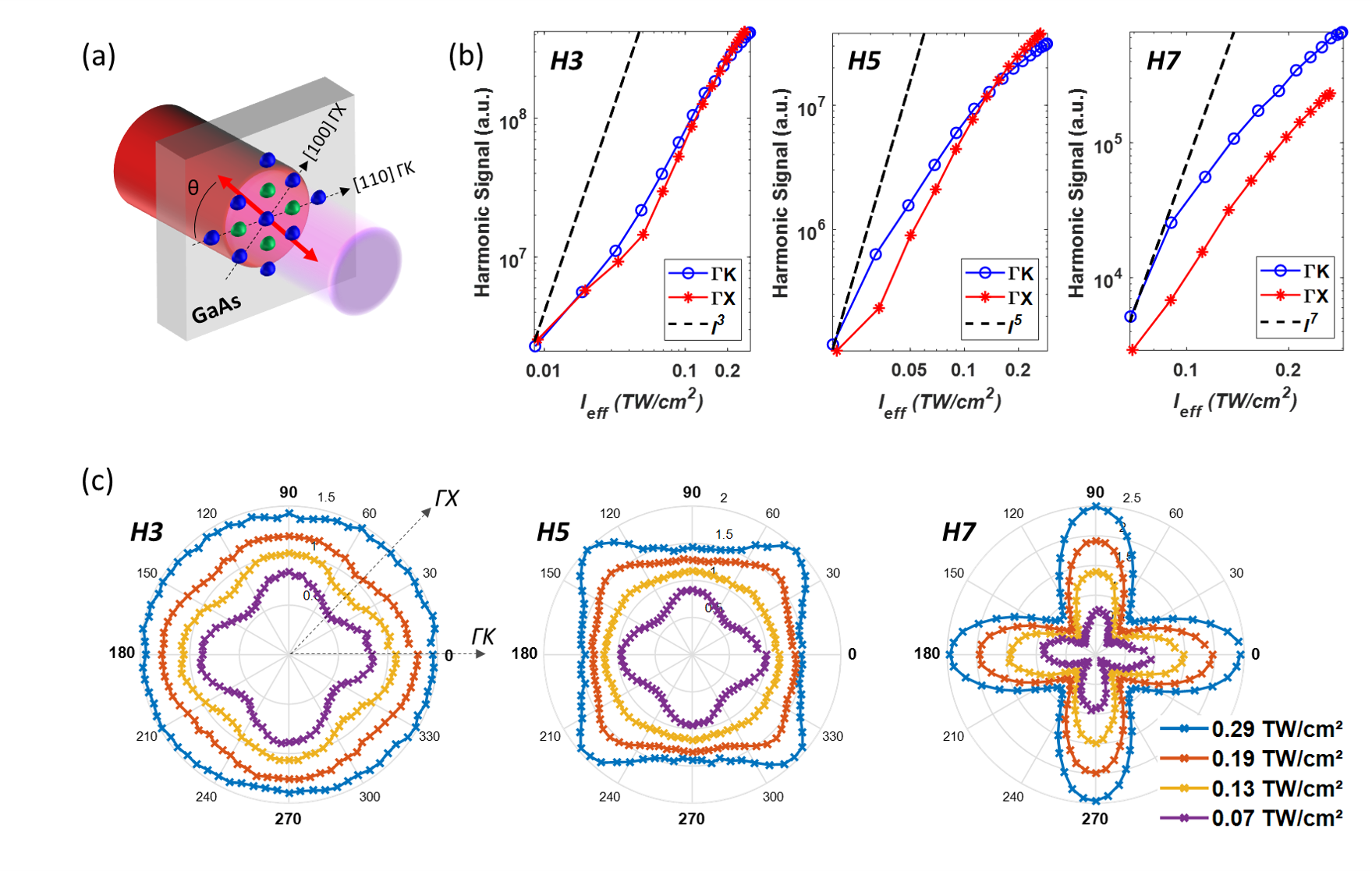}
\caption{(a) Schematic representation of the HHG experiment in transmission (t-HHG). (b) The third (H3), fifth (H5) and seventh (H7) harmonic signals emitted along the ${\Gamma K}$ and ${\Gamma X}$ axes of the GaAs crystal as a function of the effective laser intensity ${\it{I_{eff}}}$ calculated in section 3.2. The perturbative scaling is plotted in a dashed black line. (c) The polarization dependences of harmonics 3, 5 and 7 of the 2.1 µm driving laser generated from GaAs at different laser intensities. The indicated intensities are the estimated ones along the ${\Gamma K}$ direction. Note that the normalization factor of each scan is chosen only for a good visualization of the polarization dependences.}
\label{tHHG_polar scans}
\end{figure}

A schematic representation of the geometry of the t-HHG experiment, including the relative orientations of the crystal axes with the laser polarization, is shown in figure \ref{tHHG_polar scans}(a). The laser polarization is initially aligned along the ${\Gamma K}$ direction of the crystal in its (100) plane (i.e. at $\mathrm{\theta = 0^\circ}$). Being limited to the maximum pulse energy delivered by the laser, harmonics up to the $\mathrm{7^{th}}$ order (300 nm) are efficiently generated in GaAs. 
\noindent
The dependence of each harmonic yield on the laser intensity is measured along the two main crystal axes ${\Gamma K}$ and ${\Gamma X}$, and shown in a double logarithmic scale in figure \ref{tHHG_polar scans}(b). The scaling of harmonics 3 (H3), 5 (H5) and 7 (H7) clearly drifts from the perturbative scaling law (represented in black dashed lines), which verifies that the HHG process takes place in the non-perturbative regime. In addition, a remarkable behavior in the laser intensity dependence of the $\mathrm{3^{rd}}$ and $\mathrm{5^{th}}$ harmonic emissions is observed. As can be seen in figure \ref{tHHG_polar scans}(b), at low intensities the harmonic generation is more efficient along the ${\Gamma K}$ direction of the GaAs crystal (blue curves). However, at an intensity around 0.15 $\mathrm{TW/cm^2}$, an inversion in the yield of H3 and H5 between the two main axes takes place, in which the $\mathrm{3^{rd}}$ and $\mathrm{5^{th}}$ harmonic emissions become stronger along the ${\Gamma X}$ direction. To push these findings further, we carry out polarization-dependent measurements of each generated harmonic at different laser intensities. The corresponding polar plots are depicted in figure \ref{tHHG_polar scans}(c). The laser polarization is initially aligned along the ${\Gamma K}$ direction of the crystal, and the polarization is then rotated in its (100) plane. At a low laser intensity (0.07 $\mathrm{TW/cm^2}$ - violet plots), the polarization dependence of all harmonics reveals a similar fourfold symmetry. In this case, the harmonic efficiency is maximum when the laser polarization is aligned along the ${\Gamma K}$ direction of the crystal (at $\mathrm{0^{\circ}}$ and by every $\mathrm{90^{\circ}}$), and is minimum along the ${\Gamma X}$ direction ($\mathrm{45^{\circ}}$ rotation from the ${\Gamma K}$ directions). Such anisotropy is expected as GaAs has a zincblende crystal structure and thus reveals the fourfold symmetric nature of the crystal. As the harmonic order increases, for instance for H7, a stronger dependence is measured. Surprisingly, for slightly higher laser intensities, a change in the polarization dependences of the $\mathrm{3^{rd}}$ and $\mathrm{5^{th}}$ harmonic emissions is observed. For H3, as we increase the laser intensity the contrast of the anisotropic response diminishes, while for H5, its generation efficiency becomes isotropic at an effective intensity of 0.13 $\mathrm{TW/cm^2}$. For higher laser intensities, a rotation by $\mathrm{45^{\circ}}$ in the polarization dependence is observed, where the harmonic signal maximizes along the ${\Gamma X}$ crystal axis instead. This change and dependence become stronger at our maximum laser intensity of 0.29 $\mathrm{TW/cm^2}$. On the other hand, the polarization dependence of H7 at any laser intensity remains the same.

\subsection{Laser Propagation Effects}
\begin{figure}[htbp]
\centering\includegraphics[width=12cm]{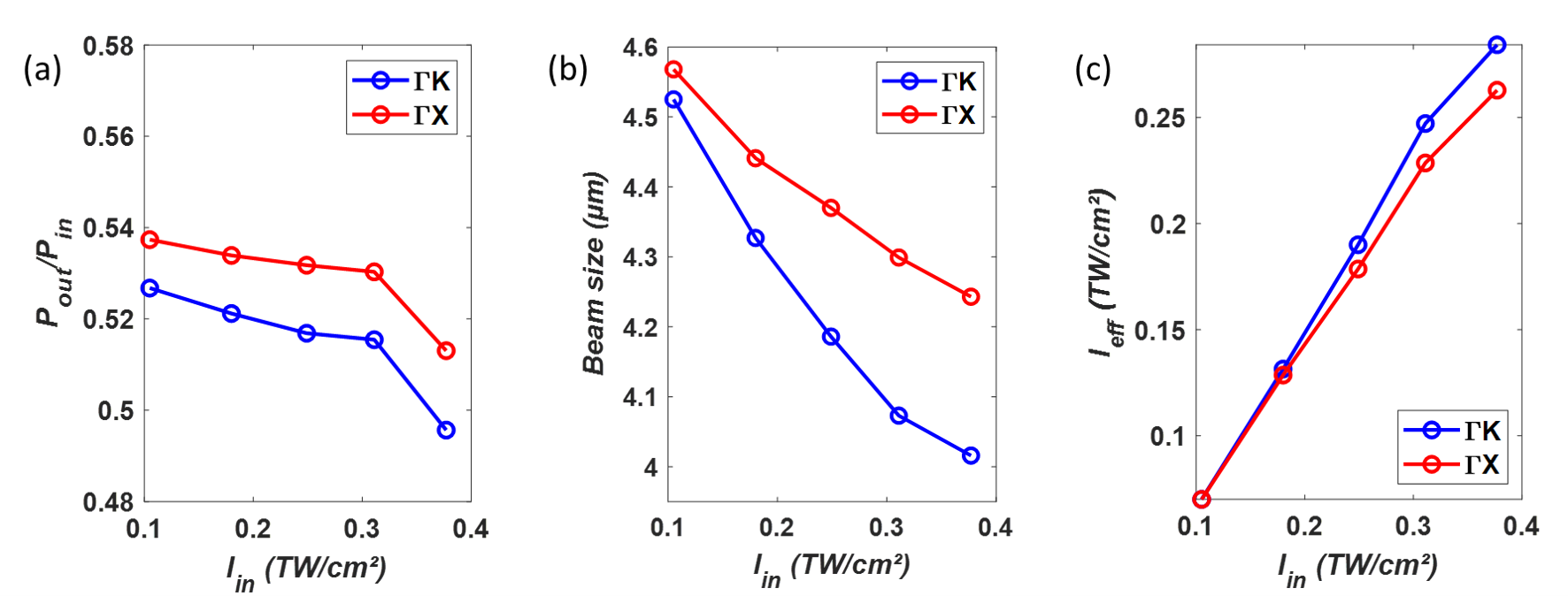}
\caption{(a) The ratio of the output power to the input power of the laser, (b) the estimated beam size at full width half maximum, and (c) the calculated effective intensity ${\it{I_{eff}}}$ close to the rear surface of the GaAs crystal versus the input laser intensity ${\it{I_{in}}}$ along the two main crystal axes ${\Gamma K}$ (in blue) and ${\Gamma X}$ (in red).} 
\label{propagation_characteristics}
\end{figure}

Given the solid state nature of the generation medium, nonlinear propagation effects can come into play for thick crystals (few tens of micrometers). This can significantly impact the laser propagation through the crystal before reaching the effective harmonic generation region \cite{xia_nonlinear_2018,Lu2019,orenstein2019}. Therefore, it is necessary to disentangle these effects from the real physical mechanism of high harmonic generation. GaAs is characterised by its high linear and nonlinear refractive indices at wavelengths around 2.1 µm, where ${n_{0}}$ = 3.38 and ${n_{2}}$ = $\mathrm{10^{-13}\,cm^{2}/W}$ \cite{hutchings_theory_1995, hurlbut_multiphoton_2007}, respectively, almost three orders of magnitude higher than those of $\mathrm{SiO_{2}}$ \cite{adair_nonlinear_1989}. Besides, the nonlinear refractive index and the multi-photon absorption coefficient at a wavelength of around 2 µm are slightly larger along the ${\Gamma K}$ than along the ${\Gamma X}$ axis of the crystal. Consequently, the Kerr effect is expected to be stronger along the ${\Gamma K}$ direction. Indeed, Xia et al. \cite{xia_nonlinear_2018} previously showed that the anisotropic multiphoton absorption in GaAs results in an increase of the harmonic emission along the ${\Gamma K}$ direction of a thick GaAs crystal compared to a thin one. In our case, the absorption of the fundamental beam along the two axes of the crystal is estimated by measuring the transmitted output average power after the GaAs sample. This includes the Fresnel transmission and the nonlinear absorption in the crystal. The measurements for different laser input intensities (as calculated from the input power and the focal spot without the sample) are reported in figure \ref{propagation_characteristics}(a).
\noindent
In agreement with previous studies, the absorption along the ${\Gamma K}$ direction of GaAs is slightly stronger than that along ${\Gamma X}$ axis. In addition, we characterize the self-focusing effect by calculating the fundamental mode size at the exit of the GaAs crystal from the direct measurement of the $\mathrm{5^{th}}$ harmonic mode. The results are displayed in figure \ref{propagation_characteristics}(b). Our measurements reveal that with increasing laser intensity the fundamental beam size decreases, and at a faster rate when the polarization is along the ${\Gamma K}$ direction than along ${\Gamma X}$. Finally, combining the output power measurements and the beam size results in a slightly higher effective intensity along the ${\Gamma K}$ axis (figure \ref{propagation_characteristics}(c)). Consequently, this would not explain the pronounced $\mathrm{3^{rd}}$ and $\mathrm{5^{th}}$ harmonic yields along the ${\Gamma X}$ direction with the increase of the laser intensity. Moreover, it is worth noting that in our tight focusing geometry, since the Rayleigh length is much shorter than the sample thickness, the overall nonlinear propagation effects are minimized. A numerical study of the beam propagation shows that it is mainly ruled by ionization-induced refraction and absorption by the plasma produced on the leading front of the pulse, on the one hand, and by the Kerr-induced self-focusing, on the other hand. For our laser parameters, the latter counterbalances plasma defocusing and leads to the reduction of the beam size observed in the experiment. In the temporal and spectral domains, a modification of the laser pulse and a spectral blue shift take place, mainly governed by photoionization of the valence band of the generation medium \cite{hussain_spectral_2021}. However, no clear difference was found between propagation along the ${\Gamma K}$ and ${\Gamma X}$.

\subsection{HHG in Reflection Geometry}
\begin{figure}[htbp]
\centering\includegraphics[width=12cm]{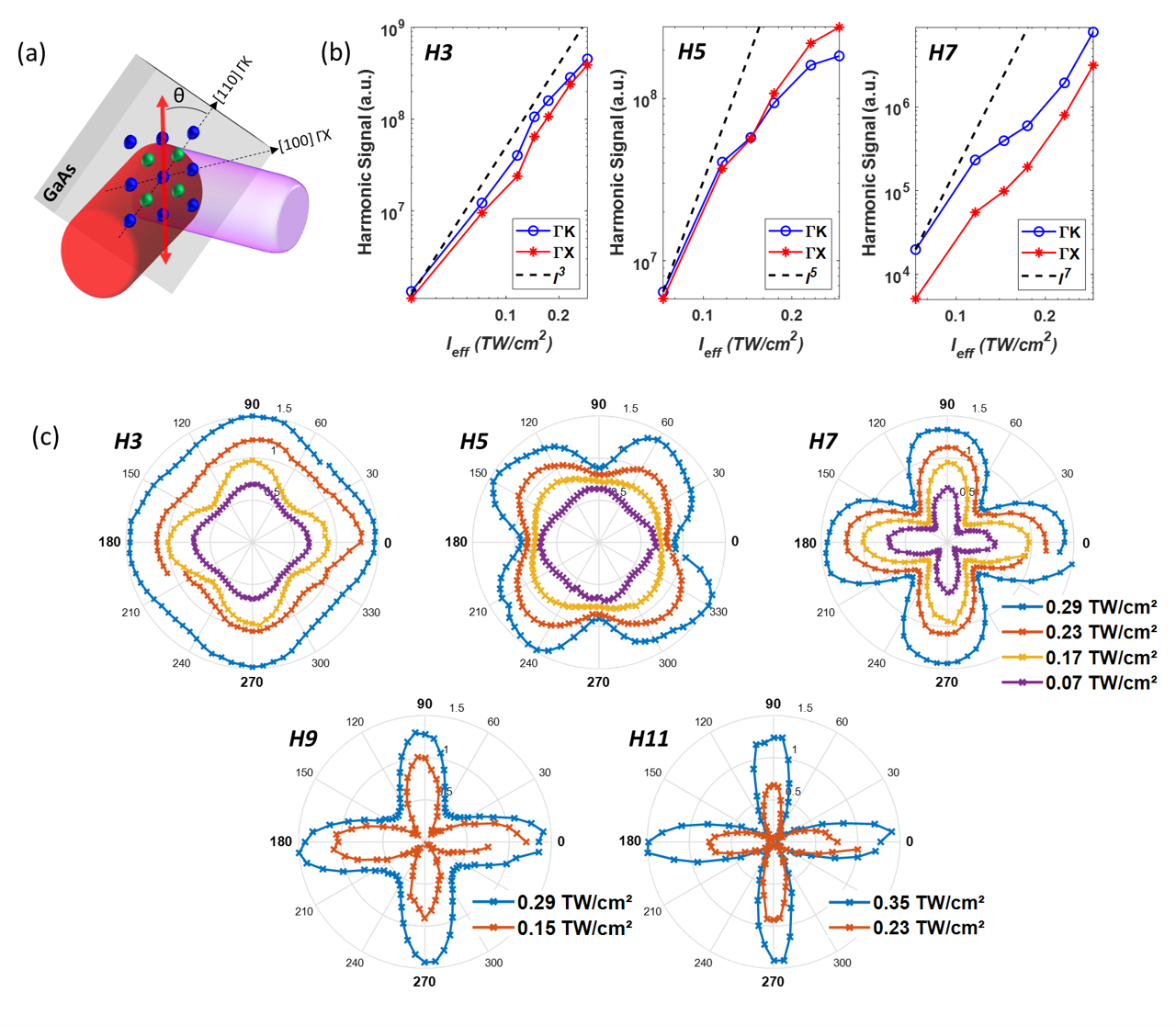}
\caption{(a) Schematic representation of the HHG experiment in reflection geometry (r-HHG). (b) The third (H3), fifth (H5) and seventh (H7) harmonic signals emitted along the ${\Gamma K}$ and ${\Gamma X}$ axes of the GaAs crystal as a function of the effective laser intensity ${\it{I_{eff}}}$. The perturbative scaling is shown by the dashed black line. (c) The polarization dependences of the harmonics H3-H11 of the 2.4 µm driving laser generated from GaAs at different laser intensities.}
\label{rHHG_polar scans}
\end{figure}

To confirm the change in the polarization dependence of low order harmonics generated in GaAs, we explore the polarization mapping of each individual harmonic generated in reflection at different laser intensities. This geometry excludes the nonlinear propagation effects that could act on the laser beam and alter the pure microscopic interaction with the generation medium \cite{vampa_observation_2018}. The geometry of the r-HHG experiment is schematized in figure \ref{rHHG_polar scans}(a). The laser is set for s-polarization, and in this case the sample is rotated about the normal to its (100) plane during the measurements. With the higher intensities provided by the OPCPA laser system, high-order harmonics are efficiently generated from GaAs, limited by our detection system to the $\mathrm{11^{th}}$ order. Damage on the sample appears above an effective intensity of 0.35 $\mathrm{TW/cm^{2}}$. The intensity scaling of harmonics 3 (H3), 5 (H5) and 7 (H7) are extracted from their polarization dependence measurements at different laser intensities (figure \ref{rHHG_polar scans}(b)). Interestingly, the inversion in the $\mathrm{5^{th}}$ harmonic yield appears again starting at a laser intensity of around 0.15 $\mathrm{TW/cm^{2}}$. Figure \ref{rHHG_polar scans}(c) reports the polarization mapping of the generated odd-order harmonics.
\noindent
Even though the wavelength used in reflection is different, similar results as those of the t-HHG experiment are obtained. Particularly, the remarkable change in the polarization dependence of the $\mathrm{3^{rd}}$ and $\mathrm{5^{th}}$ harmonic emissions as a function of the laser intensity is observed. At a low intensity of 0.07 $\mathrm{TW/cm^{2}}$, H5 signal is maximum when the laser is aligned along the ${\Gamma K}$ direction, while for relatively higher laser intensities, starting at 0.17 $\mathrm{TW/cm^{2}}$ and above, the polarization dependence of H5 rotates by ${45^{\circ}}$, i.e. the harmonic signal maximizes when the polarization is along the ${\Gamma X}$ direction. For higher order harmonics, H7, H9 and H11, the dependence on the laser polarization remains the same for different intensities. Therefore, the results of the polarization dependence of the HHG process in GaAs in reflection corroborate those obtained in transmission, and strongly confirm that the variation in the HHG anisotropy is mainly determined by the microscopic response of the crystal.

\section{Discussions}
\begin{figure}[htbp]
\centering\includegraphics[width=8cm]{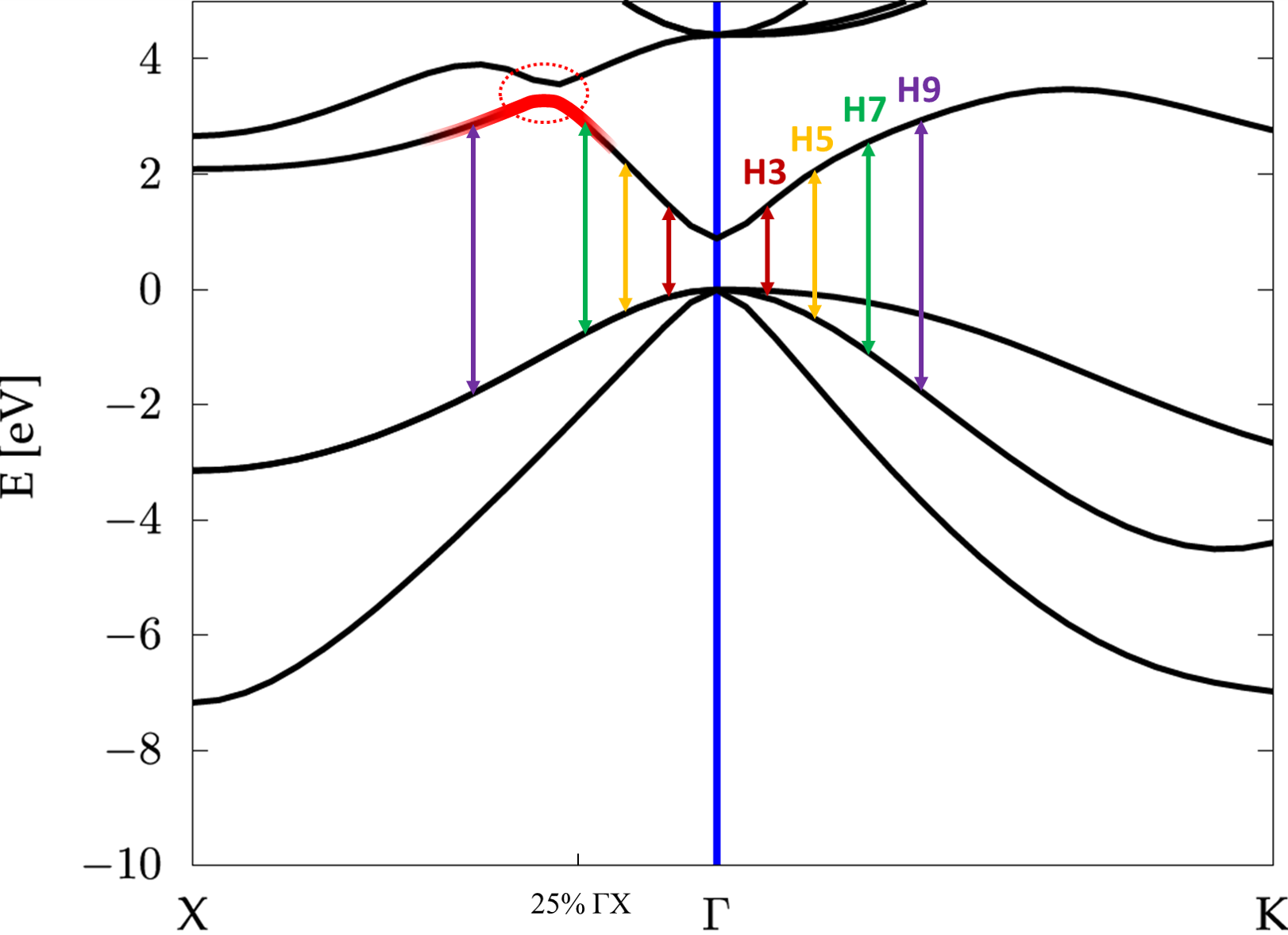}
\caption{Band structure of GaAs along the main crystal axes of interest, ${\Gamma K}$ and ${\Gamma X}$, obtained by density-functional theory (DFT) calculations \cite{NicolasOctopus, NicolasPRB2020}. The gap opening in the conduction bands' region induces a significant inflection of the first conduction band (highlighted in red), which is the source of intraband emission. Interband transitions, between the light hole and first conduction band, corresponding to the harmonic energies of the 2.1 µm driving laser are indicated.}
\label{GaAs_band structure}
\end{figure}

\begin{figure}[htbp]
\centering\includegraphics[width=13cm]{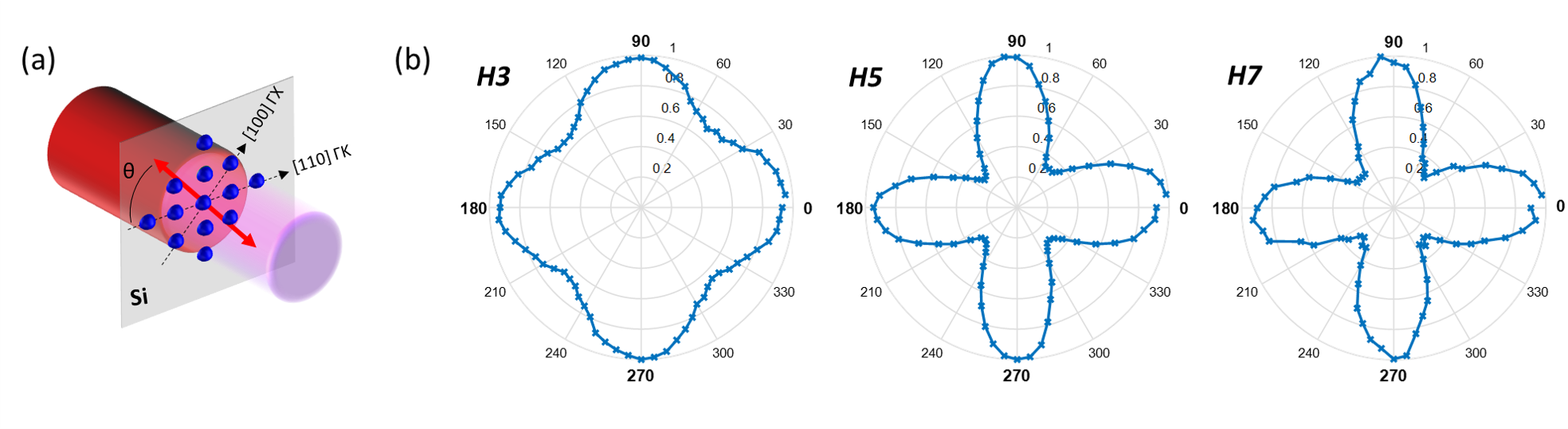}
\caption{(a) Schematic representation of the HHG experiment in a 1 µm thin silicon (100) membrane. (b) The polarization dependences of harmonics 3 (H3), 5 (H5) and 7 (H7) of the 2.1 µm driving laser generated from Si at an estimated  effective laser intensity of 0.2 $\mathrm{TW/cm^{2}}$.}
\label{tHHG_Si_experiment}
\end{figure}

\noindent 
The intensity-dependent polarization mapping observed for low-order harmonics, H3 and H5, generated from GaAs could originate from its band structure, which impacts the emission mechanisms related to the intraband and interband dynamics. A basic calculation of the electron excursion in the conduction band, using the acceleration theorem, shows that the electron can explore about ${25\%}$ of the Brillouin zone when driven at a laser intensity of 0.29 $\mathrm{TW/cm^{2}}$. Around this intensity, the electron wave packet therefore reaches a region in the band structure where a small gap exists between the first and second conduction bands of GaAs along the ${\Gamma X}$ direction \cite{chelikowsky_nonlocal_1976}, as shown in figure \ref{GaAs_band structure}.
\noindent
Around such gap, Bloch oscillations are expected to be pronounced, leading to significant nonlinear intraband current and consequently an enhancement in the low order harmonic emission \cite{ghimire_observation_2010, wu_high-harmonic_2015}. This stronger contribution of the intraband current with the increase of the laser intensity can explain the rise of the $\mathrm{3^{rd}}$ and $\mathrm{5^{th}}$ harmonic yields along the ${\Gamma X}$ direction of GaAs. However, for higher order harmonics such as harmonics 7, 9 and 11, the generation is expected to remain dominated mainly by interband transitions, being more efficient along the ${\Gamma K}$ direction \cite{vampa_theoretical_2014,yoshikawa_interband_2019}. Notice that the interband transitions for H3, H5 and H7 are allowed energetically and are relatively close to each other along the ${\Gamma X}$ direction. We would then expect that their yields evolve similarly for a dominant interband transition with the laser intensity, which is not what we observe. To support our hypothesis, we compare the polarization-dependent HHG emission from GaAs to silicon, given that both semiconductors possess the same lattice system, as well as similar valence and first conduction bands along the ${\Gamma K}$ direction, although the direct bandgap of silicon is larger than that of GaAs \cite{chelikowsky_nonlocal_1976}. However, the small additional gap present between the first and second conduction bands of GaAs along the ${\Gamma X}$ direction does not exist in silicon. Consequently, all harmonic orders present similar polarization dependences at high laser intensity, as shown in figure \ref{tHHG_Si_experiment}.
\noindent

\section{Conclusion}
In summary, we have investigated high harmonic generation from GaAs in the ultraviolet and visible spectral range, both in transmission and reflection geometries. An interesting intensity-dependent polarization mapping of low order harmonics is observed. We have successfully shown that nonlinear propagation effects are negligible in comparison to the pure HHG response, and that the change in the polarization dependence of the harmonics with the laser intensity originates from microscopic mechanisms of the generation process. The results of our work point out an interesting competition between the intraband and interband electron dynamics in the conduction band of GaAs in the strong-field regime. This would contribute to the fundamental understanding of the HHG process in semiconductors, and provide a means to differentiate between the two mechanisms involved. Finally, solid-state HHG could mark an exciting development of an all-optical band structure reconstruction technique.

\section*{Funding}
We acknowledge the financial support from the French ASTRE program through the “NanoLight” grant, the ANR (PACHA grant number ANR-17\_CE30-0008-01), as well as the support from the PETACom project (Petahertz Optoelectronics Communication) FET Open H2020 grant number 829153, OPTOLogic project (Optical Topologic Logic) FET Open H2020 grant number 899794, and TSAR project (Topological Solitons in Antiferroics) FET Open H2020 grant number 964931.

\section*{Acknowledgments}
The authors would like to truly thank Laure Lavoute, Dmitry Gaponov, Nicolas Ducros and Sébastien Février for providing the NOVAE Brevity laser.

\section*{Disclosures}
The authors declare no conflicts of interest.

\section*{Data availability}
Data underlying the results presented in this paper are not publicly available at this time but may be obtained from the authors upon reasonable request.

\end{document}